\documentclass[conference]{IEEEtran}

\usepackage{graphicx}
\usepackage{amsmath}
\usepackage{mathtools}
\usepackage[makeroom]{cancel}
\usepackage{amssymb}
\usepackage{amsfonts,amsthm,bm}
\usepackage{comment}
\usepackage{tasks}
\usepackage{float}
\usepackage{algorithm}
\usepackage{xcolor}
\usepackage[noend]{algpseudocode}
\usepackage{cite}
\usepackage{hyperref}
\usepackage{enumitem}
\usepackage{bbold}
\usepackage{multicol}

\newtheorem{remark}{Remark}

\newtheorem{assumption}{Assumption}
\newtheorem{definition}{Definition}
\IEEEoverridecommandlockouts

\begin{document}
\title{Accelerated Probabilistic State Estimation in Distribution Grids via Model Order Reduction}

% Authors
\author{

\IEEEauthorblockN{Samuel Chevalier}
\IEEEauthorblockA{Massachusetts Institute of Technology\\
Cambridge, Massachusetts\\
schev@mit.edu}

\and

\IEEEauthorblockN{Luca Schenato}
\IEEEauthorblockA{University of Padova\\
Padova, Italy\\
schenato@dei.unipd.it}\thanks{This work was supported in part by the Skoltech-MIT Next Generation grant and by the MIT Energy Initiative Seed Fund Program.}

\and

\IEEEauthorblockN{Luca Daniel}
\IEEEauthorblockA{Massachusetts Institute of Technology\\
Cambridge, Massachusetts\\
luca@mit.edu}

}

\maketitle
\begin{abstract}
This paper applies a custom model order reduction technique to the distribution grid state estimation problem. Specifically, the method targets the situation where, due to pseudo-measurement uncertainty, it is advantageous to run the state estimation solver potentially thousands of times over sampled input perturbations in order to compute probabilistic bounds on the underlying system state. This routine, termed the Accelerated Probabilistic State Estimator (APSE), efficiently searches for the solutions of sequential state estimation problems in a low dimensional subspace with a reduced order model (ROM). When a sufficiently accurate solution is not found, the APSE reverts to a conventional QR factorization-based Gauss-Newton solver. The resulting solution is then used to preform a basis expansion of the low-dimensional subspace, thus improving the reduced model solver. Simulated test results, collected from the unbalanced three-phase 8500-node distribution grid, show the resulting algorithm to be almost an order of magnitude faster than a comparable full-order Gauss-Newton solver and thus potentially fast enough for real-time use.
\end{abstract}
\IEEEpeerreviewmaketitle


% --------------- Section 1 --------------- %
\section{Introduction}
\IEEEPARstart{A}{ctive} management of distribution grids and their embedded resources is becoming a critically important task for network operators~\cite{Boardman:2020,Kong:2018}. In order to properly operate and control these distribution grids, knowledge of the network “state” is vitally important, both for system operators and for the automated controllers embedded in the network. In the seminal state estimation works by Fred Schweppe, e.g.~\cite{SchweppeI:1970}, the state of an electric power system is defined as ``the vector of the voltage magnitudes and angles at all network buses." Being a well established technology at the transmission level~\cite{Abur:2004}, state estimation is performed every few seconds, and its output allows system operators to make important decisions regarding power dispatch, voltage regulation, and stability assessment. The many potential similar advantages associated with distribution system state estimation (DSSE) are discussed in~\cite{Haughton:2013}. Despite these advantages, very few utilities have implemented real time DSSE in their systems~\cite{Boardman:2020,Primadianto:2017}, suggesting that more research must be completed before utilities choose to invest in and adopt DSSE as a tool for system management.

The challenges associated with DSSE are well documented in the literature; they most saliently include unbalanced operation, highly time varying loads, disparate measurements ($\mu$-PMU, SCADA, other AMI), and a high degree of load uncertainty. Research survey \cite{Primadianto:2017} highlights an overall lack of ``data synergy" as a primary roadblock for DSSE implementation.

%due to heterogeneous data types, measurement collection protocols, and timing coordinations, as a primary roadblock for DSSE implementation.

A variety of DSSE solvers have been proposed across the academic literature~\cite{Singh:2009,Pokhrel:2018,Schenato:2014,Dobbe:2018,Mathieu:2013,Minot:2016,Wang:2019,Bretas:2017}. Many of these approaches leverage Bayesian estimation, Kalman filtering, gross error detection, SDP relaxation, or the inclusion of regularizing pseudo-measurements. To overcome the inherent uncertainty highlighted in~\cite{Primadianto:2017}, though, advanced uncertainty quantification (UQ) approaches should be applied to DSSE, just as they have been successfully applied to the probabilistic power flow (PPF) problem~\cite{Prusty:2017}. In PPF, UQ methods, such as Polynomial Chaos Expansion~\cite{Ren:2016}, efficiently map uncertainty in the load forecast space to uncertainty in the voltage profile space. Analogously, in DSSE, UQ maps uncertainty in the ``measurement profile" space to uncertainty in the voltage profile space, as first proposed in~\cite{Lin:2011}. In massive distribution networks, with tens of thousands of state variables, sampling-based UQ approaches can require many minutes to converge if traditional DSSE solvers are employed. For real-time operation and control, this can be too slow.

Accordingly, this paper builds off the work presented in~\cite{Chevalier:2020}, where an intrusive, accelerated PPF solver is proposed. In this paper, we apply a similar projection-based model order reduction (MOR) technique to a standard set of overdetermined state estimation equations. For a given measurement sample (i.e. input), the resulting ROM is able to solve the underlying state estimation problem orders of magnitude faster than a full order solver. The associated algorithm, termed the Accelerated Probabilistic State Estimator (APSE), can then pass these solutions to any sampling-based UQ technique. The specific contributions of this paper are as follows:

\begin{enumerate}

\item We derive exact, second order expansions of the relevant distribution grid state estimation equations.

\item Leveraging an orthonormalized subspace $V$, we compress the derived expansions into a low-dimension system of equations (i.e. ROM) which can be solved rapidly.

\item To dynamically construct this subspace $V$ and initialize the reduced order model, we implement an outer-loop QR factorization-based Gauss-Newton solver.

\end{enumerate}
% The remainder of this paper is structured as follows. Section \ref{Technical Background} provides technical background on the DSSE problem, while Section \ref{sec: ROM} derives the proposed APSE algorithm. Section \ref{Num Results}, finally, provides numerical test results.
% --------------- Section 2 --------------- %
% Passivity
\section{Technical Background}\label{Technical Background}
In this section, we define a suitable network model and recall the state estimation problem. Next, we review the numerical techniques typically used to solve state estimation.

\subsection{Standard Network Model Statement}
Our state estimation methodology will be derived for a single-phase network and then extended to a three-phase network in the numerical test results section. In defining this network, we assume there is one unique bus which represents the distribution network \textit{substation}. We denote the associated graph $G({\mathcal V},\bar{\mathcal E})$, with edge set $\bar{\mathcal E}$, $|\bar{\mathcal E}|=m$, vertex set $\mathcal{V}$, $|\mathcal{V}|=n$, and signed nodal incidence matrix ${\bar E}\in{\mathbb R}^{m\times n}$. 
\begin{remark}
When defining variables which \textbf{include} the substation node, an overline will be used. For instance, ${\bar E}$ contains all network nodes, including the substation.
\end{remark}
\noindent The nodal admittance (``Y-bus") matrix takes the form
\begin{align}\label{eq: Yb}
{\bar Y}_{b}={\bar E}^{\top}Y_{l}{\bar E}+{\bar Y}_{s},
\end{align}
where $Y_{l}\in {\mathbb C}^{m\times m}$ and ${\bar Y}_{s}\in {\mathbb C}^{n\times n}$ are the  diagonal line and shunt admittance matrices, respectively. In this network, we define ${\bar{\bf V}}e^{j{\bar{\bm \theta}}}\in {\mathbb C}^{n}$ and $\bar{\bf I}e^{j\bar{\bm \phi}} \in {\mathbb C}^{n}$ as the nodal voltage and current injection phasor vectors, respectively, where ${\bar{\bf V}},\bar{\bm \theta},\bar{\bf I},\bar{\bm \phi}\in{\mathbb R}^{n}$. These vectors satisfy $\bar{\bf I}e^{j\bar{\bm \phi}} = {\bar Y}_b{\bar{\bf V}}e^{j{\bar{\bm \theta}}}$. The \textit{reduced} state vector $\bf x$ is the vector of nodal voltage magnitudes and phases at all buses \textit{except} the slack bus:
\begin{align}
{\bf x}&=\left[{\bf V}^{\top},\,{\bm{\theta}}^{\top}\right]^{\top}.
\end{align}
Assuming the slack bus voltage is known, we define the nodal active and reactive power injection function ${\bf S}({\bf x})$ as
\begin{align}
{\bf S}({\bf x})=\left[\begin{array}{c}
{\rm Re}\{\texttt{d}(\bar{{\bf V}}e^{j\bar{\bm{\theta}}})(Y_{b}\bar{{\bf V}}e^{j\bar{\bm{\theta}}})^{*}\}\\
{\rm Im}\{\texttt{d}(\bar{{\bf V}}e^{j\bar{\bm{\theta}}})(Y_{b}\bar{{\bf V}}e^{j\bar{\bm{\theta}}})^{*}\}
\end{array}\right]
\end{align}
where ${\texttt d}(\cdot)$ and $(\cdot)^*$ are the diagonalization and complex conjugation operators, respectively. Next, we define the power flow function ${\bf F}({\bf x})$. On each line, there are two potential flow functions: from the ``sending" end to the ``receiving" end, and vice versa. Without loss of generality, we assume measurement flow devices are only located on the sending end of the lines. We thus define flow function
\begin{align}
{\bf F}({\bf x})=\left[\begin{array}{c}
{\rm Re}\{\texttt{d}(\bar E_{1}\bar{{\bf V}}e^{j\bar{\bm{\theta}}})(Y_{l}\bar{E}\bar{{\bf V}}e^{j\bar{\bm{\theta}}})^{*}\}\\
{\rm Im}\{\texttt{d}(\bar E_{1}\bar{{\bf V}}e^{j\bar{\bm{\theta}}})(Y_{l}\bar{E}\bar{{\bf V}}e^{j\bar{\bm{\theta}}})^{*}\}
\end{array}\right],
\end{align}
where matrix $\bar E_1=(|\bar E|\!+\!\bar E)/2$ selects sending end voltages. For completion, we also define the voltage magnitude function ${\bf M}({\bf x})$, which simply selects voltage magnitude coordinates:
\begin{align}
{\bf M}({\bf x})={\bf V}.
\end{align}
As written, ${\bf S}({\bf x})$, ${\bf F}({\bf x})$, ${\bf M}({\bf x})$ return \textit{all} network injections, flows, and voltage magnitudes. At the physical locations where these quantities are not measured, the corresponding equations from these functions will be eventually removed.

\subsection{Power System State Estimation}
In the state estimation problem, we seek to construct the unknown state vector $\bf x$ from a set of noisy (or incomplete) measurements in order to minimize a residual cost function. In this paper, we assume there are three types of measurements\footnote{If $\mu$-PMU data is available, a phase angle measurement set $\mathcal T$ can be added to the formulation. Since $\mu$-PMUs are rare, this isn't considered here.}: voltage magnitudes, power flows, and power injections.
\begin{enumerate}
    \item $\mathcal{M}$, $|\mathcal{M}|={\mathfrak m}$, is the set of voltage \textit{magnitudes} which are measured. The associated magnitude residual function is ${\bf m}_i({\bf x})$, and $\breve{\bf m}_i=\breve{\bf V}_i$ is the $i^{\rm th}$ magnitude measurement.
    \item $\mathcal{F}$, $|\mathcal{F}|={\mathfrak f}$, is the set of active and reactive power \textit{flows} that are measured. The flow residual function is ${\bf f}_{i}({\bf x})$ and $\breve{\bf f}_i$ is the $i^{\rm th}$ power flow measurement.
    \item $\mathcal{S}$, $|\mathcal{S}|={\mathfrak s}$, is the set of active and reactive power \textit{injections} that are measured. The associated residual function is ${\bf s}_{i}({\bf x})$, and $\breve{\bf s}_i$ is the $i^{\rm th}$ injection measurement.
\end{enumerate}
Each residual function takes the form of a predictive function of the state $\bf x$ minus a measurement:
\begin{subequations}\label{eq: resid_eqs}
\begin{align}
{\bf m}_{i}({\bf x}) & = {\bf M}_{i}({\bf x})-\breve{{\bf m}}_{i}, \;i\in {\mathcal M}\label{eq: mag_func}\\
{\bf f}_{i}({\bf x}) & = {\bf F}_{i}({\bf x})-\breve{{\bf f}}_{i}, \,\;\;\;\;i\in {\mathcal F}\label{eq: flow_func}\\
{\bf s}_{i}({\bf x}) & = {\bf S}_{i}({\bf x})-\breve{{\bf s}}_{i}, \,\;\;\;\;i\in {\mathcal S}.\label{eq: inj_func}
\end{align}
\end{subequations}
We concatenate these residual functions (\ref{eq: resid_eqs}) into vector ${\bf r}({\bf x})$:
\begin{align}\label{eq: r_orig}
{\bf r}({\bf x})&=[{\bf m}({\bf x})^{\top},\,{\bf f}({\bf x})^{\top},\,{\bf s}({\bf x})^{\top}]^{\top}.
\end{align}
The typical state estimator seeks to minimize the unconstrained function $K({\bf x})$ over the field $\bf x$:
\begin{align}\label{eq: SE_opt}
\min_{{\bf x}}\;K({\bf x})=\min_{{\bf x}}\;{\tfrac{1}{2}}{\bf r}({\bf x})^{\top}\Sigma^{-1}{\bf r}({\bf x}),
\end{align}
where $\Sigma$ is the diagonal covariance matrix associated with the concatenated measurement vector $\breve{{\bf r}}=[\breve{{\bf m}}^{\top},\,\breve{{\bf f}}^{\top},\,\breve{{\bf g}}^{\top}]^{\top}$.
\begin{definition}
Vector $\breve{{\bf r}}$ is referred to as a \textbf{measurement profile}.
\end{definition}
\begin{assumption}\label{asm: nSV}
The substation voltage is known and therefore not included as an unknown variable in the state estimator.
\end{assumption}
Due to measurement redundancy, the DSSE residual equations (\ref{eq: r_orig}) are typically cast as an \textit{overdetermined} set of nonlinear equations~\cite{Singh:2009}, i.e., the associated DSSE problem (\ref{eq: SE_opt}) has a unique, generally nonzero solution. This paper targets the situation proposed in~\cite{Lin:2011}, where the DSSE problem is technically overdetermined, but a subset of the pseudo-measurements are extremely uncertain (with uniform distribution). Thus, our proposed APSE will sample from this uncertain measurement space and produce rapid DSSE system solves.
\begin{comment}
\begin{assumption}\label{asm: OD}
The DSSE equations (\ref{eq: r_orig}) are overdetermined, and (\ref{eq: SE_opt}) has a unique minimizer, but a subset of the measurements are highly uncertain with uniform distribution.
\end{assumption}
\end{comment}
\noindent Defining $p\equiv n-1$ as the number of unknown \textit{complex} voltages, measurement redundancy necessitates ${\mathfrak m}+{\mathfrak f}+{\mathfrak s}>2p$.

\subsection{Numerical State Estimation Solution Techniques}
In solving the state estimation problem (\ref{eq: SE_opt}), first order optimally condition $\partial K({\bf x})/\partial{\bf x}_i=0,\, \forall {\bf x}$
must be satisfied. Employing Gauss-Newton, we linearize $
{\bf r}({\bf x}_0+\Delta{\bf x})\approx {\bf r}({\bf x}_0)+J_{\bf r}\Delta{\bf x}$,
where ${\bf x}={\bf x}_0+\Delta{\bf x}$ and $J_{\bf r}=\frac{\partial {\bf r}({\bf x})}{\partial{\bf x}}=[
J_{{\bf m}}^{\top}, \; J_{{\bf f}}^{\top},\;  J_{{\bf s}}^{\top}]^{\top}$. Approximating,
\begin{align}
K({\bf x}) & \approx\tfrac{1}{2}\left({\bf r}_0\!+\!J_{{\bf r}}\Delta{\bf x}\right)^{\top}\Sigma^{-1}\left({\bf r}_0\!+\!J_{{\bf r}}\Delta{\bf x}\right)\equiv \tilde{K}(\Delta{\bf x}).
\end{align}
Optimality condition
${\partial}\tilde{K}(\Delta{\bf x})/{\partial\Delta{\bf x}}=0$ is satisfied by
\begin{align}\label{eq: Gx}
G({\bf x}_{0})\Delta{\bf x}=-J_{\bf r}^{\top}\Sigma^{-1}{\bf r}({\bf x}_{0}).
\end{align}
where $G({\bf x}_{0})\!=\!J_{\bf r}^{\top}\Sigma^{-1}J_{\bf r}$.
The associated iterative state estimation solution scheme becomes
\begin{align}\label{eq: Gain_mat}
{\bf x}^{(k+1)}={\bf x}^{(k)}-G({\bf x}^{(k)})^{-1}J_{\bf r}^{\top}\Sigma^{-1}{\bf r}({\bf x}^{(k)}).
\end{align}
\subsubsection*{Alternative Solution via QR Factorization}
In many applications, the ``gain" matrix $G({\bf x})$ can be poorly conditioned, so it is desirable to avoid solving the associated linear system in (\ref{eq: Gain_mat}) directly. As an alternative, we consider the over-determined and weighted linear system of equations
\begin{align}\label{eq: NSlinsys}
    \Sigma^{-\frac{1}{2}}J_{{\bf r}}\Delta{\bf x}=-\Sigma^{-\frac{1}{2}}{\bf r}({\bf x}_{0}).
\end{align}
Factoring $\Sigma^{-\frac{1}{2}}J_{{\bf r}}$ into the product of an orthogonal matrix $\mathcal Q$ and upper right triangular matrix $\mathcal R$, i.e. ${\mathcal Q}{\mathcal R}=\Sigma^{-\frac{1}{2}}J_{{\bf r}}$, the iterative scheme (\ref{eq: Gain_mat}) updates to
\begin{align}\label{eq: QR_sol}
{\bf x}^{(k+1)}={\bf x}^{(k)}-{\mathcal R}^{-1}({\mathcal Q}^{\top}\Sigma^{-\frac{1}{2}}{\bf r}({\bf x}^{(k)})).
\end{align}
\begin{comment}
In the absence of poorly conditioned numerics, (\ref{eq: Gain_mat}) and (\ref{eq: QR_sol}) take the same step and should converge to the same solution.
\end{comment}
\begin{definition}
We define the standard iterative algorithm (\ref{eq: QR_sol}) as the Gauss-Newton via QR (\textbf{GNvQR}) solution technique.
\end{definition}

\section{Model Order Reduction of the State Estimation Problem}\label{sec: ROM}
In this section, we apply the model order reduction scheme proposed in~\cite{Chevalier:2020} to the probabilistic state estimation problem.

\subsection{An Exact Expansion of the State Estimation Equations}
We seek to express the residuals (\ref{eq: resid_eqs}) as \textit{exact} second order expansions. This can be achieved for (\ref{eq: flow_func})-(\ref{eq: inj_func}) by simply moving to Cartesian voltage coordinates, where the power flow and injection functions become quadratic. We write reduced voltage vector ${\bf x}$ in Cartesian coordinates as
\begin{align}\label{eq: xc}
{\bf x}_c=\left[{\boldsymbol V}_{\rm r}^{\top},\,{\boldsymbol V}_{\rm i}^{\top}\right]^{\top}.
\end{align}
In Cartesian, the second order expansion of the voltage magnitude residual function (\ref{eq: mag_func}), will not be exact, since the expansion of $(V_{{\rm r}}^{2}+V_{{\rm i}}^2)^{\frac{1}{2}}$ has infinite terms. Therefore, we choose to update (\ref{eq: mag_func}) by taking the difference of the squares:
\begin{subequations}\label{eq: mtf}
\begin{align}
\widetilde{{\bf m}}_{i}({\bf x}_{c})&={\bf M}_{i}({\bf x})^{2}-\breve{{\bf m}}_{i}^{2}\\
&=(V_{{\rm r},i}^{2}+V_{{\rm i},i}^{2})-\breve{{\rm V}}_{i}^{2},\quad i\in{\mathcal{M}}.\label{eq: mt}
\end{align}
\end{subequations}
The effects of this minor change can be compensated for by altering the associated weights in covariance matrix $\Sigma$. 

To form the expansions of (\ref{eq: flow_func})-(\ref{eq: inj_func}) and (\ref{eq: mtf}), we consider some nominal system state, characterized by ${\bf x}_{c0}$, and some deviation from this nominal value, characterized by $\delta {\bf x}_c = {\bf x}_c - {\bf x}_{c0}$. We write the residual expansions as exact functions of this perturbation, valid $\forall \delta {\bf x}_c$:
\begin{align}
\widetilde{\bf m}(\delta{\bf x}_{c}) & ={\bf M}_{c0}^2\!+\!J_{\widetilde{\bf m},c0}\delta{\bf x}_{c}\!+\!\tfrac{1}{2}H_{\widetilde{\bf m},c}\!\left(\delta{\bf x}_{c}\!\otimes\!\delta{\bf x}_{c}\right)\!-\!\breve{{\bf m}}^2\label{eq: r1e}\\
{\bf f}(\delta{\bf x}_{c}) & ={\bf F}_{c0}\!+\!J_{{\bf f},c0}\delta{\bf x}_{c}\!+\!\tfrac{1}{2}H_{{\bf f},c}\left(\delta{\bf x}_{c}\!\otimes\!\delta{\bf x}_{c}\right)\!-\!\breve{{\bf f}}\label{eq: r2e}\\
{\bf s}(\delta{\bf x}_{c}) & ={\bf S}_{c0}\!+\!J_{{\bf s},c0}\delta{\bf x}_{c}\!+\!\tfrac{1}{2}H_{{\bf s},c}\left(\delta{\bf x}_{c}\!\otimes\!\delta{\bf x}_{c}\right)\!-\!\breve{{\bf s}},\label{eq: r3e}
\end{align}
where $\otimes$ is the Kronecker product, and ${\bf M}_{c0} \equiv {\bf M}({\bf x}_{c0})$, $J_{\widetilde{\bf m},c0} \equiv J_{\widetilde{\bf m},c}({\bf x}_{c0})$, etc. For the derivation of the Jacobian and Hessian terms, see Appendix \ref{App_Taylor}. Concatenating the function, constant, Jacobian, and Hessian terms of (\ref{eq: r1e})-(\ref{eq: r3e}), and then multiplying through by weighting matrix $\Sigma^{-\frac{1}{2}}$, we get
\begin{align}\label{eq: r}
\boldsymbol{r}(\delta{{\bf x}}_{c})=\boldsymbol{R}_{c0}+\boldsymbol{J}_{c0}\delta{{\bf x}}_{c}+\tfrac{1}{2}\boldsymbol{H}_{c}(\delta{{\bf x}}_{c}\!\otimes\!\delta{{\bf x}}_{c})-\breve{\boldsymbol r}.
\end{align}
In (\ref{eq: r}), $\boldsymbol{R}_{c0}=\Sigma^{-\frac{1}{2}}[{\bf M}_{c0}^{2\top},\,{\bf F}_{c0}^{\top},\,{\bf S}_{c0}^{\top}]^{\top}$, for example. We note that (\ref{eq: r}) contains measurement profile $\breve{\boldsymbol r}$; this will be updated (i.e. perturbed) on each sequential state estimation solve.

\subsection{Reduced Order Modeling}
We now hypothesize the existence of some subspace $V\in {\mathbb R}^{2p\times q}$, $q\ll 2p$, whose column space can accurately approximate the solution of a state estimation problem. That is, ${\bf x}_{c}\approx V\hat{{\bf x}}_c$,
where $\hat{\bf x}_c\in{\mathbb R}^q$ is a very low-dimensional vector. Due to linearity, $V\hat{{\bf x}}_{c}=V\hat{{\bf x}}_{c0}+V\delta\hat{{\bf x}}_{c}$. We now substitute $V\delta\hat{{\bf x}}_{c}\approx\delta{\bf x}_{c}$ into residual function (\ref{eq: r}):
\begin{align}\label{eq: rhi}
\!\!\!\boldsymbol{r}(\delta\hat{{\bf x}}_{c})\!=\!\boldsymbol{R}_{c0}\!+\!\boldsymbol{J}_{c0}V\delta\hat{{\bf x}}_{c}\!+\!\tfrac{1}{2}\boldsymbol{H}_{c}(V\!\otimes\! V)(\delta\hat{{\bf x}}_{c}\!\otimes\!\delta\hat{{\bf x}}_{c})\!-\!\breve{\boldsymbol r}.
\end{align}
\begin{remark}
While the state estimation equations of (\ref{eq: r_orig}) are overdetermined due to measurement redundancy, system (\ref{eq: r}) is overdetermined to a much higher degree since we have reduced input dimensionality from ${\bf x}\!\in\!{\mathbb R}^{2p}$ to $\delta\hat{\bf x}_c\!\in\!{\mathbb R}^{q}$, $q \!\ll \!2p$.
\end{remark}
To further reduce (\ref{eq: r}), we implement a projection proposed in~\cite{Chevalier:2020}, where we multiply (\ref{eq: r}) through by $\hat{\boldsymbol J}\equiv(\boldsymbol{J}_{c0}V)^\top$:
\begin{align}\label{eq: r_hat}
    \hat{\boldsymbol{r}}(\delta\hat{{\bf x}}_{c})=\hat{\boldsymbol{R}}_{c0}\!+\!\hat{\boldsymbol{G}}_{c0}\delta\hat{{\bf x}}_{c}\!+\!\tfrac{1}{2}\hat{\boldsymbol{H}}_{c}(\delta\hat{{\bf x}}_{c}\!\otimes\!\delta\hat{{\bf x}}_{c})\!-\!\hat{\breve{\boldsymbol r}},
\end{align}
whose variable definitions may be inferred. Since (\ref{eq: r_hat}) now has the same number of equations and unknowns, the reduced residual vector $\hat{\boldsymbol{r}}(\delta\hat{{\bf x}}_{c})$ can be driven to zero. The approximated Newton-like iterative solution for (\ref{eq: r_hat}) is given by
\begin{align}\label{eq: del_xGh}
\delta\hat{{\bf x}}_{c}^{(i+1)}=\delta\hat{{\bf x}}_{c}^{(i)}-\hat{\boldsymbol{G}}_{c0}^{-1}\hat{\boldsymbol{r}}(\delta\hat{{\bf x}}_{c}^{(i)}).
\end{align}
System (\ref{eq: del_xGh}) is extremely low dimensional compared to analogous system (\ref{eq: Gain_mat}), and it can therefore be iterated very rapidly. When $\hat{\bm r}\rightarrow {\bf 0}$ and (\ref{eq: del_xGh}) converges, the low-dimensional solution is converted back to the full order state vector: ${\bf x}_{c}\leftarrow {\bf x}_{c0}+V\delta\hat{{\bf x}}_{c}$. As the state estimator solves new measurement profiles, the Jacobian $\boldsymbol{J}_{c0}$ is continuously recycled, but $\hat{\boldsymbol{R}}_{c0}=(\boldsymbol{J}_{c0}V)^{\top}\boldsymbol{R}_{c0}$, $\hat{\boldsymbol{G}}_{c0}=(\boldsymbol{J}_{c0}V)^{\top}(\boldsymbol{J}_{c0}V)$, and $\hat{\boldsymbol{H}}_{c}=(\boldsymbol{J}_{c0}V)^{\top}\boldsymbol{H}_{c}(V\!\otimes\!V)$ are updated as basis $V$ expands.
\begin{definition}
We define the iterative algorithm (\ref{eq: del_xGh}) as the Reduced Model State Estimator (\textbf{RMSE}).
\end{definition}

\subsection{Dynamic Subspace Expansion}
In order to dynamically construct the orthonormalized subspace $V$, we leverage the dynamic subspace expansion approach proposed in~\cite{El-Moselhy:2010}. Our version of this procedure takes sequential \textit{state estimation} solutions ${\bf x}_c$ and projects them onto the subspace: ${\bm v} = {\bf x}_c-VV^{\top}{\bf x}_c$. If $\left\Vert {\bm v} \right\Vert\!>\!\epsilon$, then $V$ is updated:
\begin{align}\label{eq: update_V}
V={\bm [}V\;\,{\bm v}/\left\Vert {\bm v}\right\Vert{\bm ]}.
\end{align}
When $V$ is updated, the values of $\hat{\boldsymbol{R}}_{c0}$, $\hat{\boldsymbol{G}}_{c0}$, and $\hat{\boldsymbol{H}}_{c}$ are also updated. Vector $\hat{\breve{\boldsymbol r}}$ is also updated, but this changes for each new measurement profile, regardless. For an efficient update scheme, please refer to the DSE algorithm in~\cite{Chevalier:2020}.
\begin{definition}
We define the update procedure (\ref{eq: update_V}) and the associated updates of $\hat{\boldsymbol{R}}_{c0}$, $\hat{\boldsymbol{G}}_{c0}$, and $\hat{\boldsymbol{H}}_{c}$ as the Dynamic Subspace Expansion (\textbf{DSE}) routine.
\end{definition}

\subsection{Accelerated Probabilistic State Estimation}
\begin{figure}
    \centering
    \includegraphics[width=1\columnwidth]{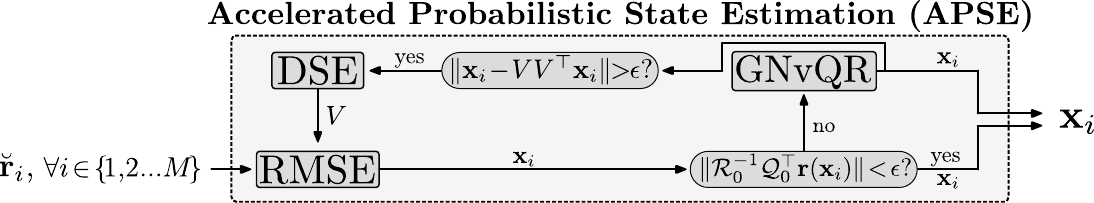}
    \caption{In the APSE algorithm, sampled measurement profiles are passed to the RMSE. If the results are of sufficient accuracy, the solver moves on to the next measurement profile. Otherwise, GNvQR solves the system. The resulting solution is then conditionally used to update basis $V$.}
    \label{fig:APSE}
\end{figure}

We now describe the full APSE procedure, which is depicted in Fig. \ref{fig:APSE}. For each new measurement profile input $\breve{\bf r}_i$, the RMSE of (\ref{eq: del_xGh}) attempts to solve the state estimation problem. If the solution does not meet the convergence criteria of the full order state estimator, then the Gauss-Newton via QR factorization algorithm of (\ref{eq: QR_sol}) is used to solve the system. Finally, if the GNvQR solution does not have a sufficiently large component is the basis, then $V$ is updated and passed back the RMSE. As basis $V$ expands, the RMSE returns solutions which are of higher and higher accuracy. For more detailed steps, Algorithm \ref{alg:APSE} provides procedural pseudocode.

Testing the quality of the RMSE solution on the full order model is a nontrivial operation. Since an optimal state estimation solution generally yields a non-zero residual, decreasing step size is often used as convergence criteria. To test if our RMSE solution meets this step size-based convergence criteria, we pre-factor a scaled Jacobian via ${\mathcal Q}_0{\mathcal R}_0=\Sigma^{-\frac{1}{2}}J_{{\bf r}}$, where, for speed, ${\mathcal Q}_0$ and ${\mathcal R}_0$ are continually recycled. If
\begin{align}\label{eq: se_test}
    \Vert\mathcal{R}_{0}^{-1}\mathcal{Q}_{0}^{\top}{\bf r}({\bf x}_{i})\Vert_{\infty}<\epsilon
\end{align}
is satisfied, then the RMSE solution is accepted. This test is shown on bottom right of Fig. \ref{fig:APSE} and line 12 of Algorithm \ref{alg:APSE}.

\begin{algorithm}
\caption{Accelerated Probabilistic State Estimator (APSE)}\label{alg:APSE}

{\small \textbf{Require:} Initial voltage solution ${{\bf x}}_{c0}$ of measurement profile $\breve{\bf r}_0$, weighted ${\boldsymbol R}_{c0}$ and weighted Jacobian ${\boldsymbol J}_{c0}$ evaluated at ${\bf x}_{c0}$, weighted Hessian ${\boldsymbol H}_{c}$, measurement profiles $\breve{\bf r}_i$ for $i=1,2,...M$

\begin{algorithmic}[1]

\Function{$[{\bf x}_1,{\bf x}_2...{\bf x}_M] \leftarrow$APSE}{$\breve{{\bf r}},{\bf x}_{c0},{\boldsymbol R}_{c0},{\boldsymbol J}_{c0},{\boldsymbol H}_c,J_{\bf r}(\cdot),\Sigma$}

\State ${\hat {\bf x}}_{c0} \leftarrow \Vert {\bf x}_{c0} \Vert$

\State $V\leftarrow {\bf x}_{c0}/{\hat {\bf x}}_{c0}$

\State $\hat{{\boldsymbol{R}}}_{c0}\leftarrow(\boldsymbol{J}_{c0}V)^{\top}{{\boldsymbol{R}}}_{c0}$

\State $\hat{\boldsymbol{G}}_{c0}\leftarrow(\boldsymbol{J}_{c0}V)^{\top}(\boldsymbol{J}_{c0}V)$

\State $\hat{\boldsymbol{H}}_{c}\leftarrow(\boldsymbol{J}_{c0}V)^{\top}{\boldsymbol{H}}_c(V\!\otimes\!V)$

\State $i\leftarrow 1$

\While{$i \le M$}

\State $\hat{\breve{\boldsymbol{r}}}_{i}\leftarrow(\boldsymbol{J}_{c0}V)^{\top}\Sigma^{-\frac{1}{2}}\breve{{\bf r}}_{i}$

\State $\delta\hat {\bf x}_c\leftarrow\,$Solve \textbf{RMSE} of (\ref{eq: del_xGh}) to convergence

\State ${\bf x}_i \leftarrow$ Cartesian-to-Polar$\{{\bf x}_{c0} + V\delta{\hat{\bf x}}_c\}$

\If{$\Vert\mathcal{R}_{0}^{-1}\mathcal{Q}_{0}^{\top}{\bf r}({\bf x}_{i})\Vert_{\infty}>$ tolerance $\epsilon_N$}

\State ${\bf x}_i \leftarrow$ Solve \textbf{GNvQR} of (\ref{eq: QR_sol}) to convergence

\State ${\bf x}_{c,i} \leftarrow $ Polar-to-Cartesian$\{{\bf x}_{i}\}$

\State Update $V,\hat{\boldsymbol{R}}_{c0},\hat{\boldsymbol{G}}_{c0},\hat{\boldsymbol{H}}_{c}$ via (\ref{eq: update_V}) and \textbf{DSE} routine

\EndIf {\bf end}

\State $i \leftarrow i + 1$

\EndWhile {\bf end}

\State \Return ${\bf x}_1,{\bf x}_2,...,{\bf x}_M$

\EndFunction

\end{algorithmic}}
\end{algorithm}

\section{Numerical Test Results}\label{Num Results}
In this section, we present test results collected from the unbalanced, three-phase IEEE 8500-node distribution network. A detailed overview of this system is provided in~\cite{Chevalier:2020}. To engender a realistic amount of measurement redundancy (5.2\% in our tests), we assume complex power flow measurements are collected on 300 medium-voltage lines, and voltage magnitude measurements are collected at 300 medium-voltage nodes. At the majority of the $\sim$2500 low-voltage loads, we assume the availability of real-time smart-meter data in the form of complex power injection measurements. 

\begin{figure}
    \centering
    \includegraphics[bb=30bp 30bp 492bp 210bp,clip,width=0.85\columnwidth]{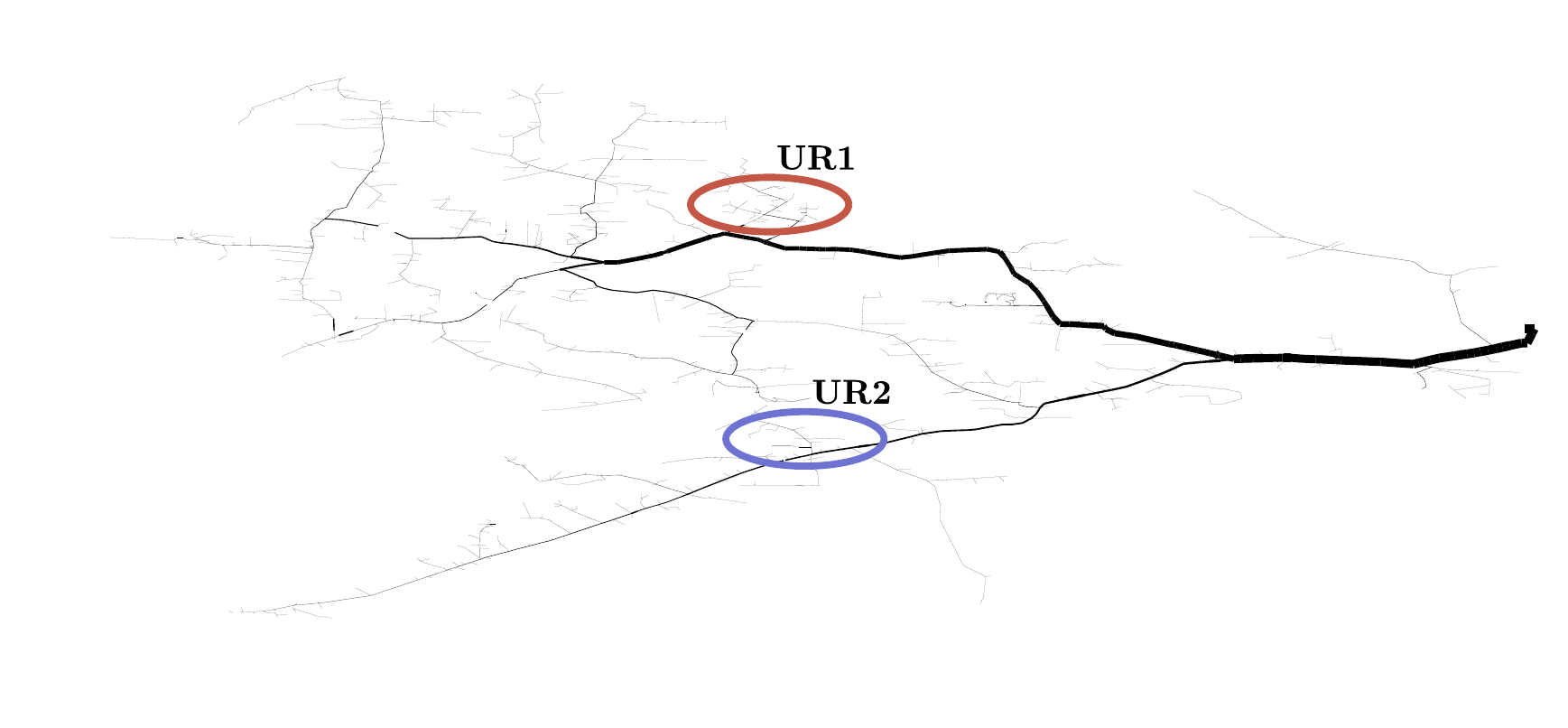}
    \caption{Map of the 8500-node network (LV distribution load buses not shown). The two measurement Uncertainty Regions (URs) are identified. In each of these URs, we identify 30 loads whose measurements are highly uncertain.}
    \label{fig:8500_Map}
\end{figure}

Next, we identify two regions in the network which contain substantial load measurement uncertainty\footnote{Such uncertainty could be caused by localized inclement weather effects, localized signal loss, or even a lack of measurement devices in some region.}. These regions are termed Uncertainty Regions (URs), and both of them are geographically identified in Fig. \ref{fig:8500_Map}. As in~\cite{Lin:2011}, we assume the load uncertainty in these URs is uniformly distributed. The upper and lower bounds of these uniform distributions are chosen as $\pm50\%$ of the historical usage mean. Next, we sample from each of these load distributions 1000 times, and we then solve 1000 instantiations of the state estimation problem (\ref{eq: SE_opt}).

\begin{figure}
    \centering
    \includegraphics[width=1\columnwidth]{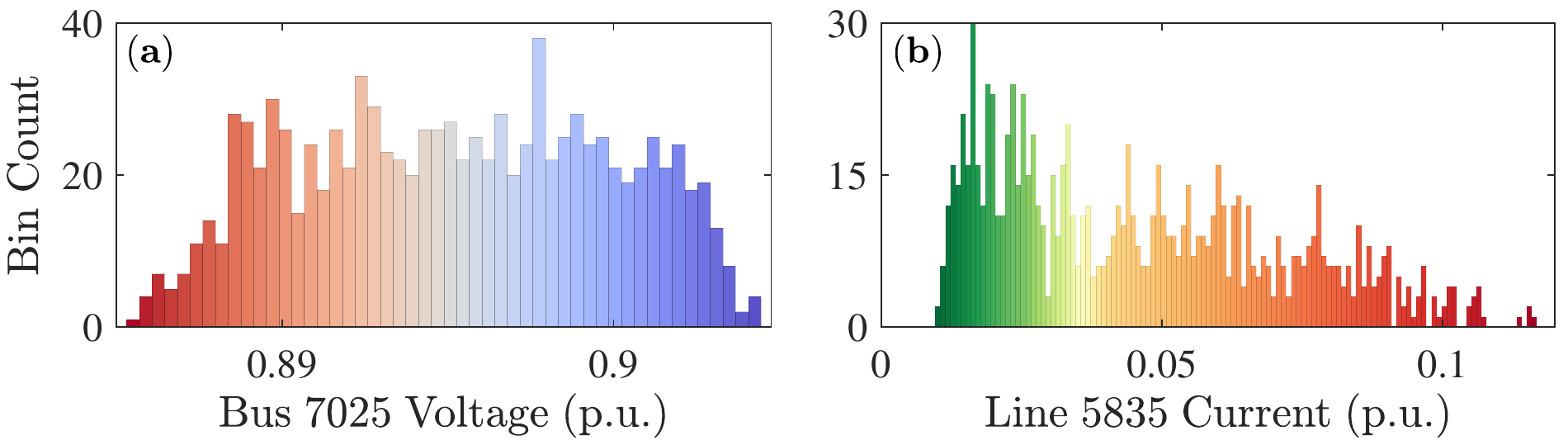}
    \caption{Shown are voltage magnitude (panel ($\bf a$)) and current magnitude (panel ($\bf b$)) histograms resulting from the 1000 solves of the 8500-node network.}
    \label{fig:VI_Dists}
\end{figure}

\begin{figure}
    \centering
    \includegraphics[width=1\columnwidth]{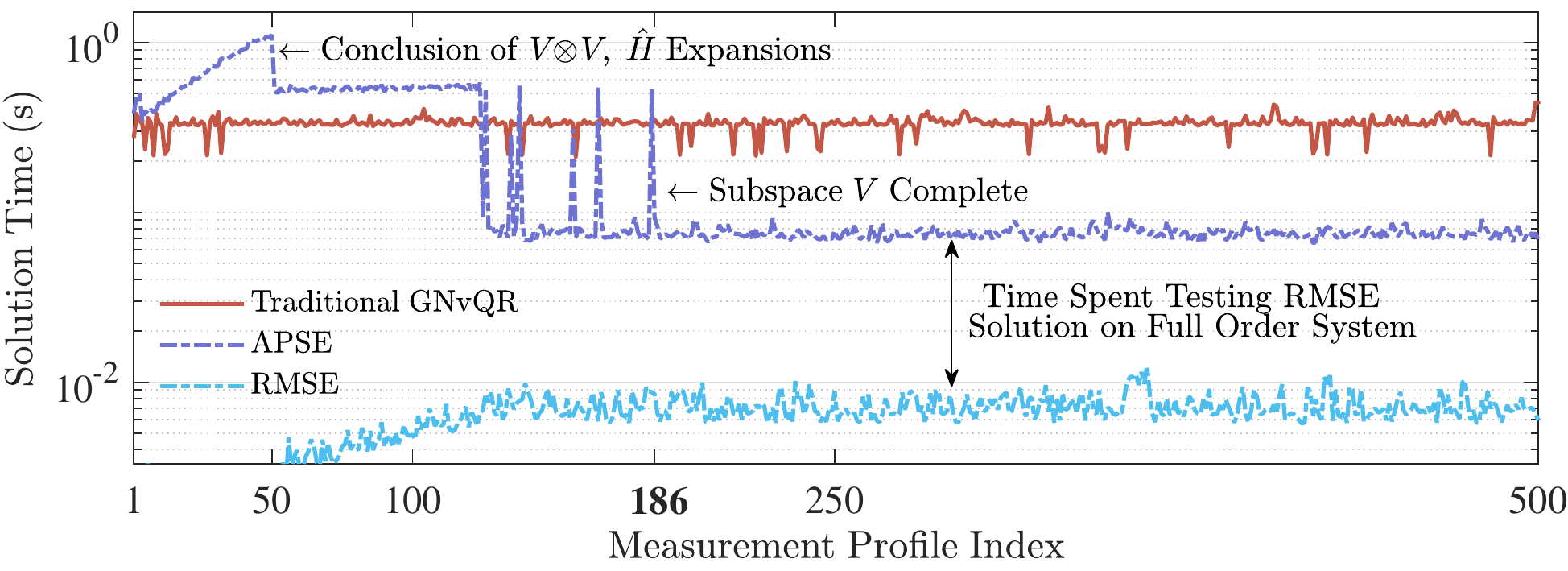}
    \caption{Timing analysis of the full order solver (GNvQR), the APSE, and the RMSE. After measurement profile 186, the APSE is 4.5 times faster than the GNvQR procedure. The amount of time spent testing the RMSE solution via (\ref{eq: se_test}) dominates the APSE and is shown by the double-arrow. As motivated in~\cite{Chevalier:2020}, the expansion of $\hat {\bm H}$ concludes after measurement profile 50.}
    \label{fig:SE_Timing}
\end{figure}

For comparison purposes, we first solve each of the resulting state estimation problems using the traditional GNvQR algorithm (\ref{eq: QR_sol}), where the the full order system had 17,062 states. Sample voltage and current results are shown in Fig. \ref{fig:VI_Dists}. Timing results are shown in Fig. \ref{fig:SE_Timing}. The average solve time of the GNvQR algorithm for each measurement profile was \textbf{0.33 seconds}, yielding a total solve time of 332 seconds.

Next, we solved the sequential state estimation problems using the APSE procedure. The timing results in Fig. \ref{fig:VI_Dists} show that once the subspace $V$ had been completed, the APSE algorithm was approximately \textbf{4.5 times faster} that the traditional GNvQR solver (at this point, the ROM had 130 internal states). Furthermore, we include the solve times of the RMSE algorithm itself (which is internal to the APSE -- see Fig. \ref{fig:APSE}). The RMSE is \textit{an order of magnitude faster} than the APSE. This is because the main bottleneck of the APSE is not the reduced model solves, but instead the testing of the reduced model solution on the full order model via (\ref{eq: se_test}).

\section{Conclusion}
This paper developed a ROM strategy for speeding up the probabilistic state estimation problem. The resulting algorithm is almost an order of magnitude faster than full order methods, but its primary computational bottleneck is the testing of the ROM solution on the full order system. Accordingly, future work will focus on developing a testing procedure which is faster than (\ref{eq: se_test}). Future work can also use deviation away from the canonical ROM operational subspace $V$ as a surrogate for anomaly or gross error detection.

\renewcommand{\baselinestretch}{.9}
\appendices
{\small
{\section{}\label{App_Taylor}}
We first consider the Jacobian terms in (\ref{eq: r1e})-(\ref{eq: r3e}). Noting that $\mathbb{V}_{\{{\rm r},{\rm i}\}}\in\mathbb{R}^{\mathfrak{m}\times p}$, the Jacobian associated with (\ref{eq: mt}) is simply
\begin{align}
    J_{\widetilde{{\bf m}},c} & \!=2\left[\mathbb{V}_{{\rm r}}\;\,\mathbb{V}_{{\rm i}}\right],\;\mathbb{V}_{\{{\rm r},{\rm i}\}_ {i,j}}=\left\{\!\!\! \begin{array}{lc}
{\boldsymbol V}_{\{{\rm r},{\rm i}\}_i},\;\;\;j=\mathcal{M}_i\\
0,\;\;\;\;\;\;\;\;\;\;\,{\rm otherwise},\label{eq: Jmc}
\end{array}\right.
\end{align}
where $\mathcal{M}_i$ refers to the $i^{\rm th}$ element of set $\mathcal M$. The flow and injection Jacobians are first given in the more familiar polar coordinates:
\begin{align}\label{eq: Js}
J_{{\bf s}} & =(\!\langle\mathtt{d}({\bf I}e^{-j\bm{\phi}})\rangle+\langle\mathtt{d}({\bf V}e^{j\bm{\theta}})\rangle N\langle Y_{b}\rangle\!)R({\bf V}e^{j\bm{\theta}})\\
J_{{\bf f}} & =(\!\langle\mathtt{d}(\hat{{\bf I}}_{l}e^{-j\hat{\bm{\phi}}_{l}})\hat{E}_{1}\rangle+\langle\mathtt{d}(\hat{E}_{1}\!{\bf V}e^{j\bm{\theta}})\rangle \hat N_{l}\langle \hat Y_{l}\hat{E}\rangle\!)R({\bf V}e^{j\bm{\theta}}).\label{eq: flow_Jac}
\end{align}
The terms $R(\cdot)$, $N$, and $\langle \cdot \rangle$ from (\ref{eq: Js}) are explicitly given in~\cite{Chevalier:2020} and are not re-stated here. Vector ${\bf I}e^{j\bm{\phi}}$ is the calculated current injections at all non-slack nodes, and vector $\hat{{\bf I}}_{l}e^{j\hat{\bm{\phi}}_{l}}$ is the calculated current flows on all lines with flow measurement devices. The other hatted terms in (\ref{eq: flow_Jac}) are equal to their non-hatted counterparts, but with the set of non-measured lines removed. The flow Jacobian (\ref{eq: flow_Jac}) represents the derivative of the flow functions (\ref{eq: flow_func}) on lines where flow measurements exist. In Cartesian, (\ref{eq: Js})-(\ref{eq: flow_Jac}) simplify to
\begin{align}
J_{{\bf s},c} & \!=\!(\!\langle\mathtt{d}(\boldsymbol{I}_{{\rm r}}\!-\!j\boldsymbol{I}_{{\rm i}})\rangle+\langle\mathtt{d}(\boldsymbol{V}_{{\rm r}}\!+\!j\boldsymbol{V}_{{\rm i}})\rangle N\langle Y_{b}\rangle\!)\label{eq: Jsc}\\
J_{{\bf f},c} & \!=\!(\!\langle\mathtt{d}(\hat{\boldsymbol{I}}_{l,{\rm r}}\!-\!j\hat{\boldsymbol{I}}_{l,{\rm i}})\hat{E}_{1}\rangle\!+\!\langle\mathtt{d}(\hat{E}_{1}\!(\boldsymbol{V}_{{\rm r}}\!+\!j\boldsymbol{V}_{{\rm i}}))\rangle {\hat N}_{l}\langle {\hat Y}_{l}\hat{E}\rangle\!).\label{eq: Jfc}
\end{align}
We observe that Jacobians (\ref{eq: Jmc}), (\ref{eq: Jsc}), (\ref{eq: Jfc}) are \textit{linear} functions of Cartesian voltage, so their associated Hessians $H_{\widetilde{\bf m},c}$, $H_{{\bf f},c}$ and $H_{{\bf s},c}$ are constant. These matrices can be found by taking the derivative of Jacobians (\ref{eq: Jmc}), (\ref{eq: Jsc}), and (\ref{eq: Jfc}) with respect to state vector ${\bf x}_c$ (\ref{eq: xc}). In the sequel, ${\bf V}_{{\rm r},i}$, for example, refers to the $i^{\rm th}$ element of the vector ${\bf V}_{{\rm r}}$ which, by definition, excludes the slack voltage. First, we define the voltage magnitude residual Hessian:
\begin{align}
& \;\;\;H_{\widetilde{{\bf m}},c} =\left[\!\!\begin{array}{ccc|ccc}
\frac{dJ_{\widetilde{{\bf m}},c}}{d{\bf V}_{{\rm r},1}} & \!\cdots\! & \frac{dJ_{\widetilde{{\bf m}},c}}{d{\bf V}_{{\rm r},p}} & \frac{dJ_{\widetilde{{\bf m}},c}}{d{\bf V}_{{\rm i},1}} & \!\cdots\! & \frac{dJ_{\widetilde{{\bf m}},c}}{d{\bf V}_{{\rm i},p}}\end{array}\!\!\right]\\
& \left.\begin{array}{c}
\frac{dJ_{\widetilde{{\bf m}},c}}{d{\bf V}_{{\rm r},i}}=2\left[\mathbb{M}\;\;{\bf 0}\right]\\
\frac{dJ_{\widetilde{{\bf m}},c}}{d{\bf V}_{{\rm i},i}}=2\left[{\bf 0}\;\;\mathbb{M}\right]
\end{array}\!\!\!\!\right\} \;\mathbb{M}_{j,k}=\left\{\!\!\! \begin{array}{lc}
1,\;\;\;\mathcal{M}_j=i,\;k=i\\
0,\;\;\;{\rm otherwise},
\end{array}\right.
\end{align}
where ${\mathfrak m}=|{\mathcal M}|$, $p=n-1$, $i\!\in\![1,...,p]$, $j\!\in\![1,...,{\mathfrak m}]$ and $k\!\in\![1,...,p]$. Next, we define the power flow residual Hessian:

\begin{align}
H_{{\bf f},c} & =\left[\begin{array}{ccc|ccc}
\frac{dJ_{{\bf f},c}}{d{\bf V}_{{\rm r},1}} & \cdots & \frac{dJ_{{\bf f},c}}{d{\bf V}_{{\rm r},p}} & \frac{dJ_{{\bf f},c}}{d{\bf V}_{{\rm i},1}} & \cdots & \frac{dJ_{{\bf f},c}}{d{\bf V}_{{\rm i},p}}\end{array}\right]\\
\tfrac{dJ_{{\bf f},c}}{d{\bf V}_{{\rm r},i}} & =\langle\mathtt{d}(\hat{Y}_{l}^{*}\hat{E}{\bf e}_{i})\hat{E}_{1}\rangle\!+\!\langle\mathtt{d}(\hat{E}_{1}{\bf e}_{i})\rangle \hat{N}_{l}\langle \hat{Y}_{l}\hat{E}\rangle\\
\tfrac{dJ_{{\bf f},c}}{d{\bf V}_{{\rm i},i}} & =\langle\mathtt{d}(-j\hat{Y}_{l}^{*}\hat{E}{\bf e}_{i})\hat{E}_{1}\rangle\!+\!\langle\mathtt{d}(j\hat{E}_{1}{\bf e}_{i})\rangle \hat{N}_{l}\langle \hat{Y}_{l}\hat{E}\rangle,
\end{align}
where $i\!\in\![1,...,p]$, and ${\bf e}_{i}\!\in\!{\mathbb R}^{p}$ is the standard unit vector. Finally, we define the power injection residual Hessian:
\begin{align}H_{{\bf s},c} & =\left[\begin{array}{ccc|ccc}
\frac{dJ_{{\bf s},c}}{d{\bf V}_{{\rm r},1}} & \cdots & \frac{dJ_{{\bf s},c}}{d{\bf V}_{{\rm r},p}} & \frac{dJ_{{\bf s},c}}{d{\bf V}_{{\rm i},1}} & \cdots & \frac{dJ_{{\bf s},c}}{d{\bf V}_{{\rm i},p}}\end{array}\right]\\
\tfrac{d{J}_{{\bf s},c}}{d{{\bf V}}_{{\rm r},i}} & =\langle\mathtt{d}(Y_{b}^{*}{\bf e}_{i})\rangle\!+\!\langle\mathtt{d}({\bf e}_{i})\rangle N\langle Y_{b}\rangle\\
\tfrac{d{J}_{{\bf s},c}}{d{{\bf V}}_{{\rm i},i}} & =\langle\mathtt{d}(-jY_{b}^{*}{\bf e}_{i})\rangle\!+\!\langle\mathtt{d}(j{\bf e}_{i})\rangle N\langle Y_{b}\rangle.
\end{align}}

\bibliographystyle{IEEEtran}
\bibliography{APSE}
\end{document}